\documentclass[aps,prd,preprintnumbers,amsmath,amssymb,nofootinbib,11pt]{revtex4}
\usepackage{eepic}
\usepackage{indentfirst}
\usepackage{mathrsfs}
\usepackage{fancyhdr}
\usepackage{ulem}
\usepackage{float}
\usepackage{graphicx}
\usepackage[
colorlinks=true, linkcolor=black, breaklinks=true, urlcolor=blue,
citecolor=green]{hyperref}
\usepackage{epstopdf}
\usepackage{bm,bbm}

\usepackage{xcolor}

\usepackage{tabularx}
\usepackage{subfigure}
\usepackage{epsfig}
\usepackage{color}
\usepackage{slashed}
\usepackage{hyperref}

\newcommand{\be}{\begin{equation}}
\newcommand{\ee}{\end{equation}}
\renewcommand{\L}{\mathscr{L}}
\newcommand{\M}{\mathscr{M}}
\newcommand{\bra}{\langle}
\newcommand{\ket}{\rangle}
\newcommand{\nn}{\nonumber}
\newcommand{\MeV}{\,\text{MeV}}
\newcommand{\GeV}{\,\text{GeV}}
\renewcommand{\vec}[1]{\mathbf{#1}}
\newcommand{\diff }{{\text{d}}}

\begin{document}
\pagestyle{plain}

\title {\boldmath Nature of the $Y(4260)$: A light-quark perspective}

\author{ Yun-Hua~Chen$^a$}
\author{ Ling-Yun~Dai$^{b}$}
\author{ Feng-Kun Guo$^{c,d}$}
\author{ Bastian Kubis$^{e}$}
\affiliation{${}^a$School of Mathematics and Physics, University
of Science and Technology Beijing, Beijing 100083, China\\
             ${}^b$School of Physics and Electronics, Hunan University, Changsha 410082, China\\
             ${}^c$CAS Key Laboratory of Theoretical Physics,
             Institute of Theoretical Physics, Chinese Academy of Sciences,
Beijing 100190, China \\
             ${}^d$School of Physical Sciences, University of Chinese Academy of
             Sciences, Beijing 100049, China\\
             ${}^e$Helmholtz-Institut f\"ur Strahlen- und Kernphysik (Theorie) and\\
             Bethe Center for Theoretical Physics,
             Universit\"at Bonn,
             53115 Bonn, Germany}

\begin{abstract}
The $Y(4260)$ has been one of the most puzzling pieces among the so-called $XYZ$ states. In this paper, we try to gain insights into the structure of the $Y(4260)$ from the light-quark perspective.
We study the dipion invariant mass spectrum of the  $e^+ e^-
\to Y(4260) \to J/\psi \pi^+\pi^-$ process and the ratio of the cross sections
${\sigma(e^+e^- \to J/\psi K^+ K^-)}/{\sigma(e^+e^- \to J/\psi \pi^+\pi^-)}$. In
particular, we consider the effects of different light-quark SU(3)
eigenstates inside the $Y(4260)$. The strong pion--pion final-state interactions
as well as the $K\bar{K}$ coupled channel in the $S$-wave
are taken into account in a model-independent way using dispersion
theory. We find that the SU(3) octet state plays a significant
role in these transitions, implying that the $Y(4260)$ contains a
large light-quark component.
Our findings suggest that the $Y(4260)$ is neither a hybrid nor a conventional charmonium state, and they are consistent with the $Y(4260)$ having a sizeable $\bar D D_1$ component which, however, is not completely dominant.
\end{abstract}

\maketitle

\newpage
\section{Introduction}

The nature of the vector charmoniumlike state $Y(4260)$ has remained controversial
since its discovery in the initial-state radiation process
$e^+e^-\to \gamma_{ISR} J/\psi\pi^+\pi^-$~\cite{Aubert:2005rm}. There is no room for
the $Y(4260)$ in the charmonium spectrum predicted in the naive quark model~\cite{Godfrey}, and the $Y(4260)$ does not show
strong couplings to ground-state open-charm decay modes~\cite{Pakhlova:2009jv},
which is unexpected for conventional vector $c\bar{c}$ states above
the $D\bar{D}$ threshold. Such peculiar properties have initiated a
lot of theoretical and experimental studies, see Refs.~\cite{Chen:2016qju,Hosaka:2016pey,Lebed:2016hpi,Esposito:2016noz,Guo:2017jvc,Ali:2017jda,Olsen:2017bmm,Karliner:2017qhf,Yuan:2018inv,Kou:2018nap,Cerri:2018ypt} for recent reviews. On the theoretical side, models have been proposed to interpret the
$Y(4260)$ as a hybrid
state~\cite{Zhu:2005hp,Close:2005iz,Kalashnikova:2008qr}, an excited
charmonium~\cite{LlanesEstrada:2005hz,Li:2009zu,Shah:2012js}, a baryonium~\cite{Qiao:2005av},
a hadrocharmonium~\cite{Dubynskiy:2008mq,Li:2013ssa}, a tetraquark
state~\cite{Maiani:2005pe,Ali:2017wsf,Wang:2018ntv}, a hadronic molecule of
$\bar{D}D_1(2420)$~\cite{Ding:2008gr,Wang:2013cya,Li:2013yla,Cleven:2013mka} or
$\omega\chi_{c0}$~\cite{Dai:2012pb}, or an interference effect~\cite{Chen:2010nv,Chen:2017uof}. On the experimental side, resonant structures with a Breit--Wigner mass ranging from $4.21$ to $4.26\GeV$
have been observed and analyzed in different channels such as
$e^+e^-\to J/\psi\pi^+\pi^-$~\cite{Aubert:2005rm,Ablikim:2016qzw},
$h_c\pi^+\pi^-$~\cite{BESIII:2016adj},
$\omega\chi_{c0}$~\cite{Ablikim:2015uix},
$X(3872)\gamma$~\cite{Ablikim:2013dyn}, $\psi'\pi^+\pi^-$~\cite{Ablikim:2017oaf},
and $D^0 D^{\ast -}\pi^++\mathrm{c.c.}$~\cite{Ablikim:2018vxx}.
The signals from all of these channels could be from the $Y(4260)$. The last one is the first observation in an open-charm channel, and the final state $D\bar D^*\pi$ is as expected from the $D\bar D_1$ hadronic molecular model~\cite{Cleven:2013mka,Qin:2016spb}.

In this work, we will study the possible light-quark components of
the $Y(4260)$ to help reveal its internal structure. We will focus
on the $\pi\pi$ invariant mass spectrum of the reaction
$e^+e^-\to Y(4260) \to J/\psi \pi\pi$, which is one of the most accurately measured channels and is the discovery channel of the $Y(4260)$. In this process, the dipion invariant mass
reaches above the $K\bar{K}$ threshold, and thus allows us to extract the
information of the light-quark SU(3) flavor-singlet and flavor-octet components.
The ratio of the cross sections ${\sigma(e^+e^- \to J/\psi K^+ K^-)}/{\sigma(e^+e^- \to J/\psi \pi^+\pi^-)}$
is relevant to the strange-quark component, and will also be taken into account. If the $Y(4260)$
contains no light quarks (as in the hybrid state or the charmonium
scenarios), the light-quark source provided by the $Y(4260)$ has
to be in the form of an SU(3) singlet state. Thus the determination
of the contributions from different SU(3) eigenstate components is
instructive to clarify the structure of the $Y(4260)$, especially in
the case if a nonzero SU(3) octet component is found to be
indispensable to reproduce the experimental data.

\begin{figure}
\centering
\includegraphics[width=\linewidth]{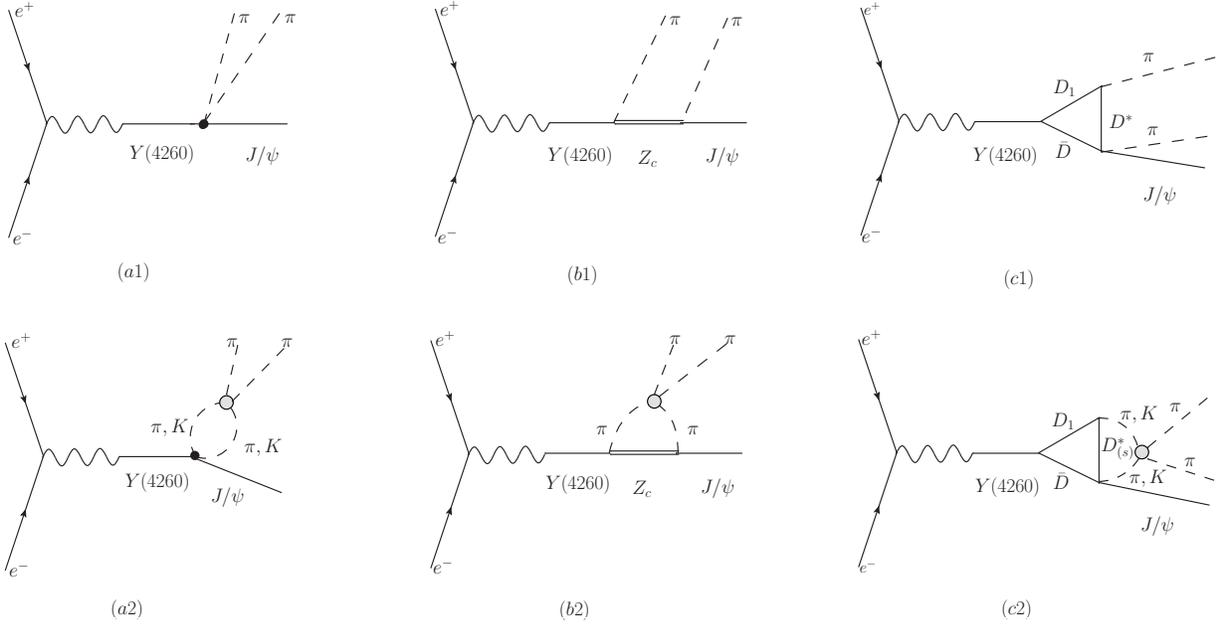}
\caption{Feynman diagrams considered for $e^+ e^- \to
Y(4260) \to J/\psi \pi \pi $. (a1) and (a2) denote
the contributions of the chiral contact $Y\psi\Phi\Phi$ terms. (b1)
and (b2) correspond to the contributions of the $Z_c$-exchange
terms. (c1) and (c2) denote the triangle diagrams. The crossed diagrams of (b1), (c1), (b2), and (c2) are not shown
explicitly. The gray blob denotes the effects of FSI.
}\label{fig.FeynmanDiagram}
\end{figure}

The conservation of parity and $C$-parity constrains the dipion system in $e^+e^-\to Y(4260) \to J/\psi \pi\pi$
to be in even partial waves. The dipion invariant mass $m_{\pi\pi}$ goes up to more than $1.1\GeV$. In this energy region, there are strong coupled-channel final-state interactions (FSIs) in the $S$-wave, which include the scalar resonances $f_0(500)$ and $f_0(980)$ and can be taken into account
model-independently using dispersion theory. Based on unitarity and
analyticity, the modified Omn\`es representation
is used in this study, where the left-hand-cut contributions are approximated
by the sum of the $Z_c(3900)$-exchange mechanism and the triangle diagrams $Y(4260) \rightarrow \bar{D}D_1(2420)\rightarrow \bar{D}D^\ast\pi (\bar{D}D_s^\ast K )\rightarrow J/\psi\pi\pi(J/\psi K\bar{K})$~\cite{Cleven:2013mka,Wang:2013hga,Albaladejo:2015lob}.\footnote{We also need
to take account of the $ Y(4260) \to J/\psi
K \bar K $  process in the coupled-channel FSI.} At low energies, the amplitude should agree with the leading chiral results,
so the
subtraction terms in the dispersion relations can be determined by matching to the chiral contact terms.
For the leading contact couplings for $Y(4260)J/\psi\pi\pi$ and $Y(4260)J/\psi K\bar K$, we construct the chiral Lagrangians in the spirit of
the chiral effective field theory ($\chi$EFT) and the heavy-quark nonrelativistic
expansion~\cite{Mannel}. The parameters are then fixed from fitting to the BESIII data.
A diagrammatic representation of all contributions is given in Fig.~\ref{fig.FeynmanDiagram}.

This paper is organized as follows. In Sec.~\ref{theor}, we describe
the theoretical framework and elaborate on the calculation of the
amplitudes as well as the dispersive treatment of the FSI. In Sec.~\ref{pheno}, we present the fit results
and discuss the light-quark components of the $Y(4260)$ and its
structure. A brief summary is given in Sec.~\ref{conclu}.

\section{Theoretical framework}\label{theor}
\subsection{Lagrangians}

In general, the $Y(4260)$ can be decomposed into
SU(3) singlet and octet components of light quarks,
\be
\label{eq.YComponents} |Y(4260)\rangle=a|V_1\rangle+b|V_8\rangle\,,
\ee
where $|V_1\rangle \equiv V_1^{\text{light}}\otimes
V^{\text{heavy}}=
\frac{1}{\sqrt{3}}(u\bar{u}+d\bar{d}+s\bar{s})\otimes
V^{\text{heavy}}$ and $|V_8\rangle \equiv
V_8^{\text{light}}\otimes
V^{\text{heavy}}=\frac{1}{\sqrt{6}}(u\bar{u}+d\bar{d}-2
s\bar{s})\otimes V^{\text{heavy}}$, and the ratio of the component
strengths $r\equiv b/a$ can be determined through fitting to the data.
Expressed in terms of a $3\times3$ matrix in the SU(3) flavor space, it is written as
\begin{equation}
 \frac{a}{\sqrt{3}} V_1 \cdot \mathbbm{1}+\frac{b}{\sqrt{6}} V_8\cdot \text{diag} \left( 1,  1, - 2\right)    .
\end{equation}

The effective Lagrangian for the $Y(4260)J/\psi\pi\pi$ and
$Y(4260)J/\psi K\bar{K}$ contact couplings, at leading order in the chiral expansion
and respecting the heavy-quark spin symmetry,
reads~\cite{Mannel,Chen2016,Chen:2016mjn}
\begin{equation}\label{LagrangianYpsipipi}
\L_{Y\psi\Phi\Phi} = g_1\bra V_{1}^\alpha J^\dag_\alpha \ket \bra u_\mu
u^\mu\ket +h_1\bra V_{1}^{\alpha} J^\dag_\alpha \ket \bra u_\mu u_\nu\ket
v^\mu v^\nu +g_8\bra  J^\dag_\alpha \ket \bra V_{8}^{\alpha} u_\mu u^\mu\ket
+h_8\bra J^\dag_\alpha \ket \bra V_{8}^{\alpha} u_\mu u_\nu\ket v^\mu v^\nu
+\mathrm{H.c.}\,,
\end{equation}
where $\langle\ldots\rangle$ denotes the trace in the SU(3) flavor space, $J= (\psi/\sqrt{3}) \cdot \mathbbm{1}$, and
$v^\mu=(1,\vec{0})$ is the velocity of the heavy quark. The lightest pseudoscalar mesons,
being the pseudo-Goldstone bosons from the spontaneous breaking of chiral symmetry, can
be filled nonlinearly into
\begin{equation}
u_\mu = i \left( u^\dagger\partial_\mu u\, -\, u \partial_\mu
u^\dagger\right) \,, \qquad
u = \exp \Big( \frac{i\Phi}{\sqrt{2}F} \Big)\,,
\end{equation}
with the Goldstone fields
\begin{align}
\Phi &=
 \begin{pmatrix}
   {\frac{1}{\sqrt{2}}\pi ^0 +\frac{1}{\sqrt{6}}\eta _8 } & {\pi^+ } & {K^+ }  \\
   {\pi^- } & {-\frac{1}{\sqrt{2}}\pi ^0 +\frac{1}{\sqrt{6}}\eta _8} & {K^0 }  \\
   { K^-} & {\bar{K}^0 } & {-\frac{2}{\sqrt{6}}\eta_8 }  \\
 \end{pmatrix} . \label{eq:u-phi-def}
\end{align}
Here $F$ is the pion decay constant in the chiral limit, and we take the physical value $92.1\MeV$ for it.

We need to define the $Z_c Y(4260)\pi$ and the $Z_c J/\psi
\pi$ interacting Lagrangians to calculate the contribution of the
intermediate $Z_c$ states, namely $Y(4260) \to Z_c\pi \to J/\psi \pi\pi$.
Note that there is no hint so far for the existence of a hidden-charm strange partner of the $Z_c$ state~\cite{Shen:2014gdm}.
We thus parametrize the $Z_c$ states in a matrix as
\begin{equation} \label{eq:Z-field}
Z^i_{c}=
 \left( {\begin{array}{*{3}c}
   \frac{1}{\sqrt{2}}Z^{0i}_{c} & Z^{+i}_{c} & 0  \\
   Z^{-i}_{c} & -\frac{1}{\sqrt{2}}Z^{0i}_{c} & 0 \\
   0 & 0 & 0
\end{array}} \right)\,.
\end{equation}
The leading-order Lagrangians are~\cite{Guo2011}
\begin{align}
\L_{Z_c Y\pi} = C_{Z_c Y\pi}  Y^i \bra {Z^i_{c}}^\dagger u_\mu \ket v^\mu
+\mathrm{H.c.} \,, \notag\\
\L_{Z_c\psi\pi} = C_{Z_c \psi\pi} \psi^i \bra {Z^i_{c}}^\dagger u_\mu \ket
v^\mu +\mathrm{H.c.} \,, \label{LagrangianZc}
\end{align}
which give the $S$-wave pionic vertices proportional to the pion energy.
Note that the SU(3) singlet and octet
components of the $Y(4260)$ are not distinguishable in the $Z_c Y(4260)\pi$ interaction, as the strange-quark component is irrelevant here.

In order to calculate the triangle diagrams $Y(4260) \rightarrow \bar{D}D_1(2420)\rightarrow \bar{D}D^\ast\pi (\bar{D}D_s^\ast K )\rightarrow J/\psi\pi\pi(J/\psi K\bar{K})$,\footnote{Here and in the following, $\bar D D_1$ always means the negative $C$-parity combination of $\bar D D_1$ and $ D \bar D_1$.}
we need the Lagrangians for the coupling of the $Y(4260)$ to $\bar{D}D_1$ as well as the couplings of the $D_1$ to $D^\ast\pi$ and $D_s^\ast K$~\cite{Colangelo:2005gb,Wang:2013cya,Guo:2013nza},
\begin{align}\nonumber
\L_{YD_1D}&=\frac{y}{\sqrt{2}}Y^i\left(\bar{D}_a^\dag
D_{1a}^{i\dag}-\bar{D}_{1a}^{i\dag} D_a^\dag\right)+{\rm H.c.} \,, \\
 \L_{D_1D^\ast P}&=i\frac{h^\prime}{F}
\big[3D_{1a}^i(\partial^i\partial^j\Phi_{ab})D^{*
j\dag}_{b}-D_{1a}^i(\partial^{j}\partial^j\Phi_{ab})D_{b}^{* i\dag}+ ... \big]+{\rm H.c.} \,,
\label{LagrangianD1}
\end{align}
where $P$ denotes the pseudoscalar meson $\pi$ or $K$.
We also need the Lagrangian for the $J/\psi D^\ast D \pi$ and $J/\psi D_s^\ast D K$ vertices, which at leading order in heavy-meson chiral perturbation theory
is~\cite{Mehen2013}
\begin{equation}\label{LagrangianJpsiDstarDpi}
\L_{\psi D^\ast D P}= \frac{g_{\psi P}}{2} \bra \psi\bar{H}_a^\dagger H_b^\dagger  \ket u_{ab}^0\,,
\end{equation}
where the charm mesons are collected in
$H_a=\vec{V}_a \cdot \boldsymbol{\sigma}+P_a$ with
$P_a(V_a)=(D^{(*)0},D^{(*)+},D_s^{(*)+})$, and
$\bar{H}_a=- \bar{\vec{V}}_a \cdot \boldsymbol{\sigma}+\bar{P}_a$
with
$\bar{P}_a(\bar{V}_a)=(\bar{D}^{(*)0},D^{(*)-},D_s^{(*)-})$~\cite{Mehen2008}.

The gauge-invariant $\gamma^\ast(\mu)$ and $Y(4260) (\nu)$ two-point coupling is given by
\be
iV_{\gamma^{\ast\mu}Y^\nu}=2i(g^{\mu\nu}p^2-p^\mu
p^\nu)c_\gamma \,,
\ee
where $p$ is the momentum of the virtual photon $\gamma^\ast.$

\subsection{Amplitudes of \boldmath{$ Y(4260) \to J/\psi PP $} processes}

First we consider the decay amplitude of $ Y(4260)(p_a) \to
J/\psi(p_b) P(p_c)P(p_d) $, which is described in terms of the
Mandelstam variables
\begin{align}
s &= (p_c+p_d)^2 , \qquad
t_P=(p_a-p_c)^2\,, \qquad u_P=(p_a-p_d)^2\,,\nn\\
3s_{0P}&\equiv s+t_P+u_P=
 M_{Y}^2+M_{\psi}^2+2m_P^2  \,.
\end{align}
The variables $t_P$ and $u_P$ can
be expressed in terms of $s$ and the scattering angle $\theta$
according to
\begin{align}
t_P &= \frac{1}{2} \left[3s_{0P}-s+\kappa_P(s)\cos\theta \right]\,,&
u_P &= \frac{1}{2} \left[3s_{0P}-s-\kappa_P(s)\cos\theta \right]\,, \nn\\
\kappa_P(s) &\equiv \sigma_P
\lambda^{1/2}\big(M_{Y}^2,M_{\psi}^2,s\big) \,, & \sigma_P &\equiv
\sqrt{1-\frac{4m_P^2}{s}} \,, \label{eq:tu}
\end{align}
where $\theta$ is defined as the angle between the positive
pseudoscalar meson and the $Y(4260)$ in the rest frame of the $PP$ system, and
$\lambda(a,b,c)=a^2+b^2+c^2-2(ab+ac+bc)$ is the K\"all\'en triangle function. We define $\vec{q}$ as the
3-momentum of final $J/\psi$ in the rest frame of the $Y(4260)$ with
\be \label{eq:q} |\vec{q}|=\frac{1}{2M_{Y}}
\lambda^{1/2}\big(M_{Y}^2,M_{\psi}^2,s\big) \,. \ee

For the $Y(4260) \to J/\psi \pi^+ \pi^-$ process, since the
crossed-channel exchanged $Z_c$ and $DD^\ast$ can be on-shell, the left-hand cut (l.h.c.)
produced intersects and overlaps with the right-hand cut (r.h.c.).
Implementing the modified Omn\`es solution method to obtain the
amplitude including FSI relies on the ability to separate the
amplitude into two parts having either l.h.c.\ or r.h.c.\ only. A way of separating the two has been proposed in Ref.~\cite{Moussallam:2013una}, using the
spectral representation of the resonance propagator as well as a
consistent application of the $i\epsilon$ prescription for the
energy variables.\footnote{As discussed in Ref.~\cite{Schmid:1967ojm}, the l.h.c.\ is in fact in the unphysical Riemann sheet. The proper $i\epsilon$ helps to locate the l.h.c.\ in the right position so that it does not overlap with the r.h.c.\ in the physical Riemann sheet.} Similarly we use the spectral
representations of the $Z_c$ propagator and the $D_1$ propagator~\cite{Moussallam:2013una},
\begin{align}\label{eq.SpectralPropagator}
\widetilde{BW}_R(x)=\frac{1}{\pi}\int_{x^{\text{thr}}_R}^\infty \diff x^\prime
\frac{\textrm{Im} [BW_R(x^\prime)]}{x^\prime-x}\,,
\end{align}
where
$BW_R(x^\prime)=(M_{R}^2-x^\prime-iM_{R}\Gamma_{R}(x^\prime))^{-1}$, and $R$ denotes $Z_c$ or $D_1$.
The off-shell-width effects of
the broad intermediate resonances could play a role in the process discussed~\cite{Cleven:2013mka,Qin:2016spb}, and
we construct the energy-dependent widths
for the broad vector resonances. Taking into account that the $Z_c J/\psi \pi$ vertex is in an $S$-wave and proportional to the energy of the pion, and the $D_1 \to D^\ast\pi$ decays in a $D$-wave, the energy-dependent widths of $Z_c$ and $D_1$ read
\begin{align} \label{widthi}
\Gamma_{Z_c}(s)&=\Gamma_{Z_c} \frac{E_{\psi\pi}^2(s)}{E_{\psi\pi}^2(M_{Z_c}^2)} \frac{k_{\psi\pi}(s)M_{Z_c}}{k_{\psi\pi}(M_{Z_c}^2)\sqrt{s}}\,,  \nonumber\\
\Gamma_{D_1}(s)&=\Gamma_{D_1} \frac{k_{D^\ast\pi}^5(s)M_{D_1}}{k_{D^\ast\pi}^5(M_{D_1}^2)\sqrt{s}}\,,
\end{align}
where
$k_{QP}(s)=\lambda^{{1}/{2}}(M_Q^2,m_P^2,s)/(2\sqrt{s})$ is the magnitude of the three-vector momentum of the pion, and $E_{QP}(s)=\sqrt{m_\pi^2+k_{QP}^2(s)}.$
The thresholds in Eq.~\eqref{eq.SpectralPropagator} are $x_{D_1}^{\text{thr}}=(M_D+m_\pi)^2$ and $x_{Z_c}^{\text{thr}}=(M_\psi+m_\pi)^2$, respectively.\footnote{In this paper we aim at describing the dipion invariant mass spectrum. The $Z_c$ enters only through providing parts of the l.h.c.. In this case, we can neglect the subtlety due to the closeness of the $Z_c$ mass to the $D\bar D^*$ threshold in the spectral function. On the contrary, if we want to fit to the $Z_c$ line shape, such an effect has to be taken into account properly, see Refs.~\cite{Hanhart:2015cua,Albaladejo:2015lob,Guo:2016bjq,Pilloni:2016obd}.} Notice that the integration convolves with other parts of the amplitude.  Now the $Z_c$-exchange amplitude reads
\begin{align}\label{eq.MZc}
\hat{M}^{Z_c,\pi}(s,\cos\theta)&=\frac{2}{F^2}\sqrt{M_{Y}M_{\psi}}M_{Z_c}C_{Y\psi}p_c^0
p_d^0\bigg(\widetilde{BW}_{Z_c}(t)+\widetilde{BW}_{Z_c}(u)\bigg)\nonumber\\
&=\sum_{l=0}^\infty \hat{M}_l^{Z_c,\pi}(s)P_l(\cos\theta)\,,
\end{align}
where $C_{Y\psi}^{Z_c}\equiv C_{Z_c Y\pi} C_{Z_c \psi\pi}$ is the product
of the coupling constants for the exchange of the $Z_c$. The amplitude has been partial-wave decomposed, and
$P_l(\cos\theta)$ are the standard Legendre polynomials. Parity and
$C$-parity conservation (or isospin conservation combined with Bose symmetry) require the pion pair to be in even angular momentum partial waves. We only take into account the $S$-
and $D$-wave components in this study, neglecting the effects of
higher partial waves. Explicitly, the projections of $S$- and $D$-waves
of the $Z_c$-exchange amplitude read
\begin{align}\label{eq.M0Zc}
\hat{M}_0^{Z_c,\pi}(s)&=-\frac{2 \sqrt{M_Y M_\psi}M_{Z_c}}{\pi F^2
\kappa_\pi(s)}C_{Y\psi}\int_{x_{Z_c}^{\text{thr}}}^\infty \diff x^\prime
\frac{M_{Z_c}\Gamma_{Z_c}(x^\prime)}{(x^\prime-M_{Z_c}^2)^2+M_{Z_c}^2
{\Gamma_{Z_c}^2(x^\prime)}}\Big\{
\big(s+|\vec{q}|^2\big)Q_0(y(s,x^\prime))\nonumber\\
&\quad -|\vec{q}|^2\sigma_\pi^2\big[y^2(s,x^\prime)
Q_0(y(s,x^\prime))-y(s,x^\prime)\big] \Big\},
\end{align}
and
\begin{align}\label{eq.M2Zc}
\hat{M}_2^{Z_c,\pi}(s)&=-\frac{5 \sqrt{M_Y M_\psi}M_{Z_c}}{\pi F^2
\kappa_\pi(s)}C_{Y\psi}\int_{x_{Z_c}^{\text{thr}}}^\infty \diff x^\prime
\frac{M_{Z_c}\Gamma_{Z_c}(x^\prime)}{(x^\prime-M_{Z_c}^2)^2+M_{Z_c}^2
{\Gamma_{Z_c}^2(x^\prime)}}\Big\{ \big[s+|\vec{q}|^2-|\vec{q}|^2\sigma_\pi^2
y^2(s,x^\prime)\big] \nonumber\\&\quad
\times\big[(3y^2(s,x^\prime)-1)Q_0(y(s,x^\prime))-3y(s,x^\prime)\big]
\Big\}\,,
\end{align}
respectively, where $y(s,x^\prime)\equiv {(3s_0-s-2x^\prime)}/{\kappa_\pi(s)}$,
and $Q_0(y)$ is the Legendre function of the second kind,
\begin{equation}\label{eq.Q0}
Q_0(y)=\frac{1}{2}\int_{-1}^1 \frac{\diff z}{y-z}P_0(z)
 = \frac{1}{2}\log \frac{y+1}{y-1}\,.
\end{equation}
Notice that the analytic continuation of $Q_0(y)$ should be taken into account since the
$Z_c$ can be on-shell in the physical region. There are two finite
branch points in $Q_0(y(s,x^\prime))$,
\be
s_{\pm}(x^\prime)=\frac{1}{2x^\prime}\Big\{(M_Y^2+M_\psi^2)(m_\pi^2+x^\prime)-M_Y^2 M_\psi^2-(x^\prime-m_\pi^2)^2\pm \lambda^{{1}/{2}}(M_Y^2,x^\prime,m_\pi^2)\lambda^{{1}/{2}}(M_\psi^2,x^\prime,m_\pi^2)   \Big\}\,.
\ee
In the range of $s_- < s <s_+$, the
argument of the logarithm in Eq.~\eqref{eq.Q0} becomes negative, and the continuation reads~\cite{Barton:1961ms,Bronzan:1963mby,Kambor:1995yc}
\begin{equation}\label{eq.Q0Continuation}
Q_0(y)
 = \frac{1}{2}\log \Big|\frac{y+1}{y-1}\Big|+i\frac{\pi}{2}\,.
\end{equation}

Now we briefly discuss the calculation of the triangle diagrams.
We only keep the terms proportional to
$\bm{\epsilon}_{Y}\cdot \bm{\epsilon}_{\psi}$, and omit the remaining terms proportional to contractions of
momenta with the polarization vectors, which are suppressed in the heavy-quark nonrelativistic expansion~\cite{Chen:2016mjn}.
Explicitly, the partial-wave projections of the triangle amplitude for the $ Y(4260) \to J/\psi \pi\pi (J/\psi K \bar{K}) $ process read
\begin{align}\label{eq.MlLoop}
\hat{M}_l^{\text{loop},\pi(K)}(s)&=\frac{2l+1}{2}\frac{ \sqrt{M_Y M_\psi}M_{D_1}M_{D}M_{D_{(s)}^\ast}}{4\pi F^2
}C_{Y\psi}^{\text{loop}}\int_{-1}^1 \diff \cos\theta P_l(\cos\theta) \int_{x_{D_1}^{\text{thr}}}^\infty \diff x^\prime \text{Im}[BW_{D_1}(x')]
\nonumber\\
&\times  \int  \frac{\diff^d l}{(2\pi)^d}\bigg\{\frac{i |\vec{p_d}|^2 p_c^0}{(l^2-x^\prime+i\epsilon)\big[(p_a-l)^2-M_D^2+i\epsilon\big]\big[(l-p_d)^2-M_{D_{(s)}^\ast}^2+i\epsilon\big]}
\nonumber\\& \qquad +\frac{i |\vec{p_c}|^2p_d^0}{(l^2-x^\prime+i\epsilon)\big[(p_a-l)^2-M_D^2+i\epsilon\big]\big[(l-p_c)^2-M_{D_{(s)}^\ast}^2+i\epsilon\big]}\bigg\}\,,
\end{align}
where $C_{Y\psi}^{\text{loop}}\equiv y h^\prime g_{\psi P}$ is the product of the coupling constants for the triangle diagrams.

For the chiral contact terms, using the Lagrangians in
Eq.~\eqref{LagrangianYpsipipi}, we have
\begin{align}
M^{\chi,\pi}(s,\cos\theta)&=-\frac{4}{F^2}\sqrt{M_{Y}M_{\psi}}\bigg[\Big(g_1+\frac{g_8}{\sqrt{2}}\Big)p_c\cdot
p_d +\Big(h_1+\frac{h_8}{\sqrt{2}}\Big)p_c^0 p_d^0 \bigg]\,, \notag\\
\label{eq.ContactPi+KRaw}
M^{\chi,K}(s,\cos\theta)&=-\frac{4}{F^2}\sqrt{M_{Y}M_{\psi}}\bigg[\Big(g_1-\frac{g_8}{2\sqrt{2}}\Big)p_c\cdot
p_d +\Big(h_1-\frac{h_8}{2\sqrt{2}}\Big)p_c^0 p_d^0 \bigg]\,.
\end{align}
The projections of the $S$- and $D$-waves of the chiral contact terms are
given by
\begin{align}
M_0^{\chi,\pi}(s)&=-\frac{2}{F^2}\sqrt{M_{Y}M_{\psi}}
\bigg\{\Big(g_1+\frac{g_8}{\sqrt{2}}\Big) \left(s-2m_\pi^2 \right)
+\frac{1}{2}\Big(h_1+\frac{h_8}{\sqrt{2}}\Big) \bigg[s+\vec{q}^2\Big(1
-\frac{\sigma_\pi^2}{3} \Big)\bigg]\bigg\}\,, \notag\\
M_0^{\chi,K}(s)&=-\frac{2}{F^2}\sqrt{M_{Y}M_{\psi}}
\bigg\{\Big(g_1-\frac{g_8}{2\sqrt{2}}\Big) \left(s-2m_K^2 \right)
+\frac{1}{2}\Big(h_1-\frac{h_8}{2\sqrt{2}}\Big) \bigg[s+\vec{q}^2\Big(1
-\frac{\sigma_K^2}{3} \Big)\bigg]\bigg\}\,, \notag\\
\label{eq.M0+2Pi+Kchiral}
M_2^{\chi,\pi}(s)&=\frac{2}{3
F^2}\sqrt{M_{Y}M_{\psi}}\,\Big(h_1+\frac{h_8}{\sqrt{2}}\Big)
|\vec{q}|^2\sigma_\pi^2\,.
\end{align}
For the $D$-wave, where the $\pi\pi$ scattering is almost elastic
in the energy range considered here, we only
give the amplitude of the process involving pions.

\subsection{Final-state interactions with a dispersive approach, Omn\`es solution }

There are strong FSI in the $\pi\pi$ system in particular in the
isospin-0 $S$-wave, which can be taken into account
model-independently using dispersion theory. Since the invariant
mass of the pion pair reaches above the $K\bar{K}$ threshold, we
will consider the coupled-channel ($\pi\pi$ and $K\bar
K$) FSI for the dominant $S$-wave component, while for the $D$-wave
only the single-channel ($\pi\pi$) FSI will be considered.

For $Y(4260) \to J/\psi \pi^+ \pi^-$, the
partial-wave expansion of the amplitude including FSI reads
\be
\M^\text{full}(s,\cos\theta) = \sum_{l=0}^{\infty}
\left[M_l^\pi(s)+\hat{M}_l^\pi(s)\right] P_l(\cos\theta)\,,
\label{eq.PartialWaveFullAmplitude}\ee
where $M_l^\pi(s)$ contains
the r.h.c.\ part and accounts for the $s$-channel rescattering,
and the ``hat function'' $\hat{M}_l^\pi(s)$ represents the l.h.c., contributed by the crossed-channel pole terms or the open-flavor loop effects.
In this study, we approximate the l.h.c.\ by the sum of the $Z_c$-exchange diagram and the triangle diagrams,
i.e., $\hat{M}_l^\pi(s)=\hat{M}_l^{Z_c,\pi}(s)+\hat{M}_l^{\text{loop},\pi}(s)$. The method of
approximating the l.h.c.\ in dispersion relations by including the most relevant resonance
exchanges (in the case of no loops) has been applied previously e.g.\ in
Refs.~\cite{Moussallam-gamma,KubisPlenter,ZHGuo,Kang,Dai:2014lza,Dai:2014zta,Dai:2016ytz,Chen2016}.

For the $S$-wave, we will take into account the two-channel
rescattering effects. The functions $\hat{M}_l(s)$ do not have
a r.h.c., so the two-channel unitarity condition leads to the discontinuity of the production amplitudes as
\begin{equation}\label{eq.unitarity2channel}
\textrm{disc}\, \vec{M}_0(s)=2i T_0^{0\ast}(s)\Sigma(s)
\left[\vec{M}_0(s)+\hat{\vec{M}}_0(s)\right] ,
\end{equation}
where the two-dimensional vectors $\vec{M}_0(s)$ and
$\hat{\vec{M}}_0(s)$ stand for the r.h.c.\ and the l.h.c.\
parts of both the $\pi\pi$ and the $K\bar{K}$ final states,
respectively,
 \begin{equation}
\vec{M}_0(s)=\left( {\begin{array}{*{2}c}
   {M^\pi_0(s)} \\
   {\frac{2}{\sqrt{3}}M^K_0(s)}  \\\end{array}} \right), \hspace{0.5cm}\hat{\vec{M}}_0(s)=\left( {\begin{array}{*{2}c}
   {\hat{M}^\pi_0(s)} \\
   {\frac{2}{\sqrt{3}}\hat{M}^K_0(s)}  \\
\end{array}} \right).
 \end{equation}
The two-dimensional matrices $T_0^0(s)$ and $\Sigma(s)$ are given by
\begin{equation}\label{eq.T00}
T_0^0(s)=
 \left( {\begin{array}{*{2}c}
   \frac{\eta_0^0(s)e^{2i\delta_0^0(s)}-1}{2i\sigma_\pi(s)} & |g_0^0(s)|e^{i\psi_0^0(s)}   \\
  |g_0^0(s)|e^{i\psi_0^0(s)} & \frac{\eta_0^0(s)e^{2i\left(\psi_0^0(s)-\delta_0^0(s)\right)}-1}{2i\sigma_K(s)} \\
\end{array}} \right),
\end{equation}
and $\Sigma(s)\equiv \text{diag}
\big(\sigma_\pi(s)\theta(s-4m_\pi^2),\sigma_K(s)\theta(s-4m_K^2)\big)$.
Three input functions enter the $T_0^0(s)$ matrix: the
$\pi\pi$ $S$-wave isoscalar phase shift $\delta_0^0(s)$, and the modulus and phase of the $\pi\pi \to
K\bar{K}$ $S$-wave amplitude $g_0^0(s)=|g_0^0(s)|e^{i\psi_0^0(s)}$. To estimate the uncertainty due to the
dispersive input for the $\pi\pi/K\bar{K}$ rescattering, we will use two different $T_0^0(s)$ matrices, the
Dai--Pennington (DP)~\cite{Dai:2014lza,Dai:2014zta,Dai:2016ytz} and the Bern/Orsay (BO)~\cite{Leutwyler2012,Moussallam2004} parametrizations, and compare the results.
Note that the inelasticity parameter $\eta_0^0(s)$ in Eq.~\eqref{eq.T00} is
related to the modulus $|g_0^0(s)|$ by
\begin{equation}
\eta_0^0(s)=\sqrt{1-4\sigma_\pi(s)\sigma_K(s)|g_0^0(s)|^2\theta(s-4m_K^2)}\,.
\end{equation}
These inputs are used up to $\sqrt{s_0}=1.3\GeV$, below the onset of
further inelasticities from the $4\pi$ intermediate states, where the $f_0(1370)$ and $f_0(1500)$ resonances become important that couple strongly to
$4\pi$~\cite{Tanabashi:2018oca,Ropertz:2018stk}. Above $s_0$, the phases
$\delta_0^0(s)$ and $\psi_0^0$ are guided smoothly to 2$\pi$ by
means of~\cite{Moussallam2000}
\begin{equation}
\delta(s)=2\pi+(\delta(s_0)-2\pi)\frac{2}{1+({s}/{s_0})^{3/2}}\,.
\end{equation}

\noindent The solution of the inhomogeneous coupled-channel unitarity condition in
Eq.~\eqref{eq.unitarity2channel} is given by
\begin{equation}\label{OmnesSolution2channel}
\vec{M}_0(s)=\Omega(s)\bigg\{\vec{P}^{n-1}(s)+\frac{s^n}{\pi}\int_{4m_\pi^2}^\infty
\frac{\diff
x}{x^n}\frac{\Omega^{-1}(x)T(x)\Sigma(x)\hat{\vec{M}}_0(x)}{x-s}\bigg\}
\,,
\end{equation}
where $\Omega(s)$ satisfies the homogeneous coupled-channel
unitarity relation
\begin{equation}\label{eq.unitarity2channelhomo}
\textrm{Im}\, \Omega(s)=T_0^{0\ast}(s)\Sigma(s) \Omega(s),
\hspace{1cm}  \Omega(0)=\mathbbm{1} \,,
\end{equation}
and its numerical results have been computed, e.g., in
Refs.~\cite{Leutwyler90,Moussallam2000,Hoferichter:2012wf,Daub}.

For the $D$-wave, the single-channel FSI will
be taken into account. In the elastic $\pi\pi$ rescattering region,
the partial-wave unitarity condition reads
\begin{equation}\label{eq.unitarity1channel}
\textrm{Im}\, M_2(s)= \left[M_2(s)+\hat{M}_2(s)\right]
\sin\delta_2^0(s) e^{-i\delta_2^0(s)}\,,
\end{equation}
where the phase of the isoscalar $D$-wave amplitude
$\delta_2^0$ coincides with
the $\pi\pi$ elastic phase shift, as required by
Watson's theorem~\cite{Watson1,Watson2}.
The modified Omn\`es solution of Eq.~\eqref{eq.unitarity1channel}
can be obtained as~\cite{Leutwyler96,Chen2016}
\be\label{OmnesSolution1channel}
M_2(s)=\Omega_2^0(s)\bigg\{P_2^{n-1}(s)+\frac{s^n}{\pi}\int_{4m_\pi^2}^\infty
\frac{\diff x}{x^n} \frac{\hat
M_2(x)\sin\delta_2^0(x)}{|\Omega_2^0(x)|(x-s)}\bigg\} \,, \ee
where the polynomial $P_2^{n-1}(s)$ is a subtraction function, and the
Omn\`es function is defined as~\cite{Omnes}
\begin{equation}\label{Omnesrepresentation}
\Omega_2^0(s)=\exp
\bigg\{\frac{s}{\pi}\int^\infty_{4m_\pi^2}\frac{\diff x}{x}
\frac{\delta_2^0(x)}{x-s}\bigg\}\,.
\end{equation}
We will use the result given by the Madrid--Krak\'ow group~\cite{Pelaez} for $\delta_2^0(s)$, which is smoothly continued to $\pi$ for $s\to\infty$.

In order to determine the necessary number of subtractions that
guarantees the convergence of the dispersive integrals in
Eqs.~\eqref{OmnesSolution2channel} and \eqref{OmnesSolution1channel}, we need to
investigate the high-energy behavior of the integrands. First, it is
known that for a phase shift $\delta_l^I(s)$ approaching $k\,\pi$ at
high energies, the corresponding single-channel Omn\`es function
falls asymptotically as $s^{-k}$. As a consequence, we have $\Omega_{2}^0(s) \sim 1/s$ at large $s$. Furthermore, the
coupled-channel Omn\`es function $\Omega_l^I(s)$ is found to fall
asymptotically as $1/s$ for large $s$~\cite{Moussallam2000}, provided the asymptotic condition $\sum
\delta_l^I(s) \geq 2\pi$ for $s\to\infty$, where $\sum
\delta_l^I(s)$ is the sum of the eigenphase shifts. Second, we have
checked that in the intermediate energy region of $1\GeV^2 \lesssim s
\ll M_{Y(4260)}^2$, the inhomogeneity contributed by the
$Z_c$-exchange and the triangle diagrams grows at most linearly in $s$. So we conclude
that in the dispersive representations of
Eqs.~\eqref{OmnesSolution2channel} and \eqref{OmnesSolution1channel}, three subtractions for each of them are sufficient
to make the dispersive integrals convergent. On the other hand, at
low energies the amplitudes $\vec{M}_0(s)$ and $M_2(s)$ should match to those from $\chi$EFT. Namely, in the limit of
switching off the FSI at $s=0$, $\Omega(0)=\mathbbm{1}$ and $\Omega_2^0(0)=1$,
the subtraction terms should agree well with the low-energy chiral amplitudes
given in Eq.~\eqref{eq.M0+2Pi+Kchiral}.
Therefore, for the
$S$-wave, the integral equation takes the form
\begin{equation}\label{M02channel}
\vec{M}_0(s)=\Omega(s)\bigg\{\vec{M}_0^{\chi}(s)+\frac{s^3}{\pi}\int_{4m_\pi^2}^\infty
\frac{\diff
x}{x^3}\frac{\Omega^{-1}(x)T(x)\Sigma(x)\hat{\vec{M}}_0(x)}{x-s}\bigg\}
\,,
\end{equation}
where $ \vec{M}^{\chi}_0(s)=\big(
   M_0^{\chi,\pi}(s),
   2/\sqrt{3}\,M_0^{\chi,K}(s)
   \big)^{T}$, while
for the $D$-wave, it can be written as
\be\label{M21channel}
M_2^\pi(s)=\Omega_2^0(s)\bigg\{M_2^{\chi,\pi}(s)+\frac{s^3}{\pi}\int_{4m_\pi^2}^\infty
\frac{\diff x}{x^3} \frac{\hat
M_2^\pi(x)\sin\delta_2^0(x)}{|\Omega_2^0(x)|(x-s)}\bigg\} \,.
\ee

The amplitude for $Y(4260) \to J/\psi \pi^+\pi^-$
can be expressed in terms of the ingredients discussed above as
\begin{equation}
M^{\text{decay}}(s,\cos\theta) =
M_0^\pi(s)+\hat{M}_0^\pi(s)+\left[M_2^\pi(s)+\hat{M}_2^\pi(s)\right]
P_2(\cos\theta)\,.\label{eq.DecayAmplitude}
\end{equation}
The polarization-averaged modulus-square of the $e^+e^- \to Y(4260)
\to J/\psi \pi^+\pi^-$ amplitude can be written as
\begin{equation}
|\bar{M}(E^2,s,\cos\theta)|^2 = \frac{4\pi\alpha
c_\gamma^2|M^{\text{decay}}(s,\cos\theta)|^2}{3|E^2-M_Y^2+iM_Y\Gamma_Y|^2
M_\psi^2}\left[  8 M_\psi^2 E^2+(s-E^2-M_\psi^2)^2
\right],\label{eq.eetoJpsipipiAmplitudeSquar}
\end{equation}
where $E$ is the center-of-mass energy of the
$e^+e^-$ collisions, and we set the $\gamma^\ast Y(4260)$ coupling
constant $c_\gamma$ to 1 since it can be absorbed into the overall
normalization when we fit to the event distributions. Here we use the energy-independent width for the
$Y(4260)$, and the values of the $Y(4260)$ mass and width are taken as $4222\MeV$ and $44.1\MeV$, respectively, which are the central values of the BESIII fit in Ref.~\cite{Ablikim:2016qzw}. We also have tried to allow the
mass and width to float freely, and found
that the fit quality changes only slightly.
At last, the $\pi\pi$ invariant mass distribution of $e^+e^- \to J/\psi \pi^+\pi^-$ reads
\begin{equation}
\frac{\diff\sigma}{\diff m_{\pi\pi}} =\int_{-1}^1
\frac{|\bar{M}(E^2,s,\cos\theta)|^2
|\vec{k_3^\ast}||\vec{k_5}|}{128\pi^3 |\vec{k_1}|E^2}\diff
\cos\theta\,,\label{eq.pipimassdistribution}
\end{equation}
where $\vec{k_1}$ and $\vec{k_5}$ denote the 3-momenta of $e^\pm$ and $J/\psi$ in the
center-of-mass frame, respectively, and $\vec{k_3^\ast}$ is the 3-momenta of $\pi^\pm$ in the rest frame of the
$\pi\pi$ system. They are given as
\begin{equation}
|\vec{k_1}|=\frac{E}{2}\,, \quad
|\vec{k_3^\ast}|=\frac{1}{2}\sqrt{s-4m_\pi^2}\,, \quad
|\vec{k_5}|=\frac{1}{2E} \lambda^{1/2}\big(E^2,s,M_\psi^2\big) \,.
\end{equation}
For $e^+e^- \to Y(4260) \to J/\psi K^+ K^-$, the relevant Feynman diagrams can be obtained by
replacing all external pions by kaons in
Fig.~\ref{fig.FeynmanDiagram} (for (c1), the exchanged $D^*$ needs to be replaced by $D_s^*$ ), but without diagram~(b1) due to the
absence of the $Z_c \psi K$ vertex.
Most ingredients of the amplitude
of $e^+e^- \to Y(4260) \to J/\psi K^+ K^-$ have
been given in the above.

\section{Phenomenological discussion}\label{pheno}

\subsection{Characteristics of singlet and octet contributions}

\begin{figure}
\centering
\includegraphics[width=\linewidth]{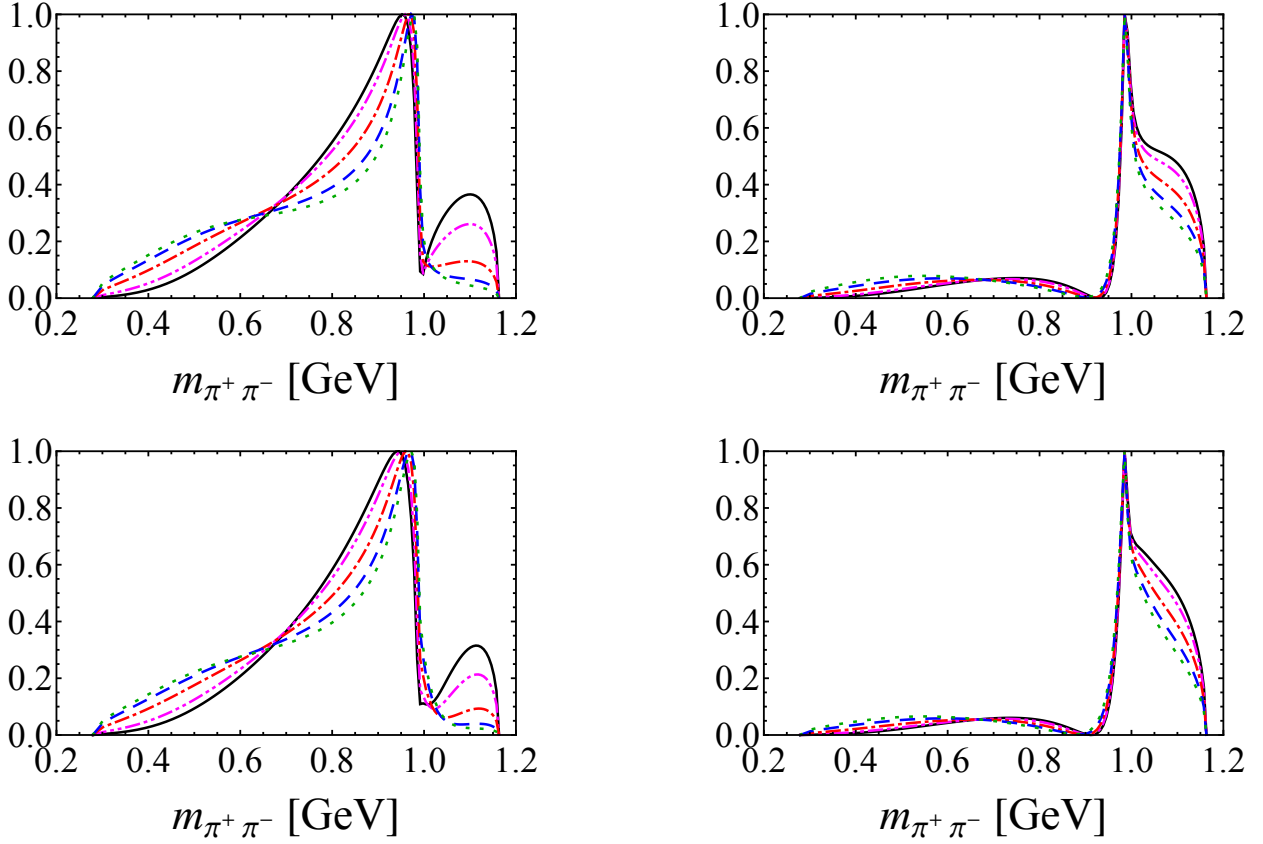}
\caption{The shapes of the $\pi\pi$ invariant mass
spectra contributed from the singlet (left) and octet (right)
chiral contact terms using the DP (top) or the BO (bottom)  parametrizations. The black solid, magenta dash-dot-dotted, red dot-dashed,
blue dashed, and green dotted lines correspond to the contributions
with $h_i/g_i$ ($i=1,8$) fixed at 0.1, 0.3, 1, 3, and 10, respectively. For the normalizations we set the highest point to be 1 for each group. }\label{fig.Mpipi_hiovergi}
\end{figure}

The two pions in the final state must come from light-flavor sources. It is instructive to discuss what would be expected for the dipion invariant mass distributions produced from pure SU(3) flavor singlet and octet sources, which are proportional to $(\bar u u + \bar d d + \bar s s)/\sqrt{3}$ and $(\bar u u + \bar d d - 2\bar s s)/\sqrt{6}$, respectively, without considering the left-hand-cut contribution. It is well known that the nonstrange and strange scalar pion form factors, $\langle 0 |(\bar u u+\bar d d)|\pi^+\pi^-\rangle$ and $\langle 0 |s\bar s|\pi^+\pi^-\rangle$, behave very differently. The former has a broad bump around $0.5\GeV$, and has a narrow dip at around $1\GeV$, while the latter has a narrow peak at around $1\GeV$. The narrow structures are manifestations of the scalar meson $f_0(980)$, which couples differently to the nonstrange and strange sources~\cite{Daub:2012mu,Daub}.  It is therefore natural to expect that the SU(3) singlet and octet pion scalar form factors should also be dramatically different.

To demonstrate the characteristic structures  in the dipion mass spectrum from the singlet and octet sources for the current problem, we need to take into account the energy dependence in the chiral contact terms. Their contributions are separately shown with varying $h_i/g_i$ in Fig.~\ref{fig.Mpipi_hiovergi}.
We consider a large range for the ratio $h_i/g_i$ ($i=1,8$). The black solid, magenta dash-dot-dotted, red dot-dashed,
blue dashed, and green dotted curves in the figure correspond to the ratio taking values of 0.1, 0.3, 1, 3, and 10, respectively. For an easy comparison, the maxima of the curves in each plot are normalized to 1.
One observes that the basic characteristic structures of both the singlet and octet spectra are stable against the variation of $h_i/g_i$: the singlet spectra display a broad bump below $1\GeV$, and around $1\GeV$ there is a dip for $h_1/g_1\lesssim 1$; the octet spectra have little contribution below $0.9\GeV$, and show a sharp peak around $1\GeV$, corresponding to the $f_0(980)$.
It is also worthwhile to notice that both of them have different behaviors from both the nonstrange and the strange pion scalar form factors.
Therefore, one expects that precise measurements of the dipion invariant mass distributions can provide valuable information about the light-quark content of the source, considering the $J/\psi$ to be a SU(3) flavor singlet.

\subsection{Fitting to the BESIII data}

In this work we perform fits taking into account the experimental data sets of the $\pi\pi$
invariant mass distributions of $e^+e^- \to J/\psi
\pi^+\pi^-$ and the ratios of the cross sections ${\sigma(e^+e^- \to J/\psi
K^+ K^-)}/{\sigma(e^+e^- \to J/\psi \pi^+\pi^-)}$ measured at two energy points
$E=4.23\GeV$ and $E=4.26\GeV$ by the BESIII
Collaboration~\cite{Collaboration:2017njt,Ablikim:2018epj}.
\begin{figure}
\centering
\includegraphics[width=\linewidth]{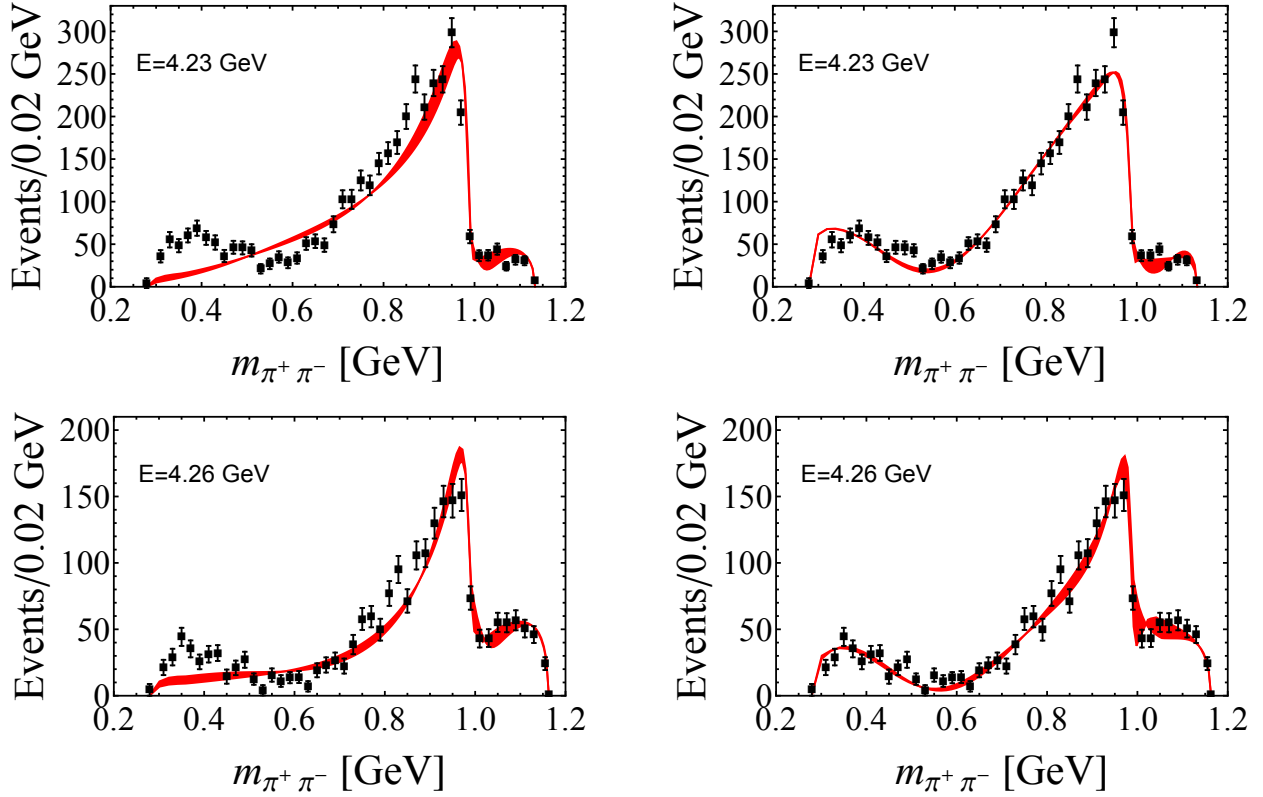}
\caption{Fit results of the $\pi\pi$ invariant mass spectra in
$e^+e^- \to J/\psi \pi^+\pi^-$ for Fits~Ia (top left),
Ib (top right), IIa (bottom left), and IIb (bottom right).
The borders of the bands represent our best fit results using two different $T_0^0(s)$ matrices.
The background-subtracted and efficiency-corrected experimental data are taken from Ref.~\cite{Collaboration:2017njt}.
}\label{fig.Mpipi}
\end{figure}
As in Refs.~\cite{Collaboration:2017njt,Pilloni:2016obd}, we regard the measurements at $E=4.23\GeV$ and $E=4.26\GeV$ as independent, and thus the coupling constants are allowed to be
different in the fits of these two data sets.
For the normalization factor for each dataset,
we choose to absorb it into the coupling constants.
There are six free parameters in our fits: $g_{1,8}$, $h_{1,8}$, $C_{Y\psi}^{Z_c}$, and $C_{Y\psi}^{\text{loop}}$. The parameters $g_1$ and $h_1$
correspond to the low-energy constants in the $Y\psi\Phi\Phi$ Lagrangian in Eq.~\eqref{LagrangianYpsipipi} for the SU(3) singlet component of the $Y(4260)$, $g_8$ and $h_8$ are the corresponding parameters for the SU(3) octet component. $C_{Y\psi}^{Z_c}$ and $C_{Y\psi}^{\text{loop}}$ are related to the
$Z_c$-exchange contribution\footnote{
The parameter $C_{Y\psi}^{Z_c}$, as a product of the $YZ_c\pi$ and $Z_c\psi\pi$ couplings, is related to the partial widths of the $Y\to Z_c\pi$ and $Z_c\to J/\psi\pi$. In principle, it can be determined from a thorough analysis of the $Z_c$ and $Y$ line shapes; such an analysis that takes into account the $\pi\pi$ FSI is not available yet. Thus, here we make a compromise by focusing on the $\pi\pi$ distribution and taking $C_{Y\psi}^{Z_c}$ as a free parameter.} and triangle-diagram contribution, respectively.
To illustrate the effect of the SU(3) octet
component, we perform two fits for each data set (Fits~Ia
and Ib for $E=4.23\GeV$, and Fits~IIa and IIb for
$E=4.26\GeV$). To be specific, in Fits~Ia and IIa
we only consider the SU(3) singlet component, the
$Z_c$-exchange terms, and the triangle diagrams, while in Fits~Ib and IIb,
the SU(3) octet components are taken into account in addition.
The coupled-channel FSI is considered in all the fits.

\begin{table}
\caption{\label{table-ratioes}  Experimental and theoretical
values  for the cross sections ratios ${\sigma(e^+e^- \to J/\psi K^+ K^-)}/{\sigma(e^+e^- \to J/\psi\pi^+\pi^-)}\times 10^{2}$. The experimental data are taken
from Ref.~\cite{Ablikim:2018epj}. The theoretical results are obtained with two different $T_0^0(s)$ matrices (DP vs.\ BO).}
\renewcommand{\arraystretch}{1.2}
\begin{center}
\begin{tabular}{l|ccccc}
\toprule
&Experiment &           Fit~Ia, DP &   Fit~Ib, DP&   Fit~Ia, BO &   Fit~Ib, BO   \\
\hline
 $\frac{{\sigma( J/\psi
K^+ K^-)}}{{\sigma( J/\psi
\pi^+\pi^-)}} \times 10^{2}, E=4.23\GeV$ &  $6.44\pm 1.15$ & $7.82\pm 0.83$& $7.75\pm 1.10$& $5.88\pm 0.82$& $2.83\pm 1.05$ \\
\toprule
&Experiment &           Fit~IIa, DP &   Fit~IIb, DP&   Fit~IIa, BO &   Fit~IIb, BO   \\
\hline
 $\frac{{\sigma( J/\psi
K^+ K^-)}}{{\sigma( J/\psi
\pi^+\pi^-)}} \times 10^{2} , E=4.26\GeV$ &  $4.99\pm 1.10$ & $4.46\pm 0.82$& $4.67\pm 0.98$& $5.37\pm 1.03$& $5.38\pm 0.82$   \\
\botrule
\end{tabular}
\end{center}
\renewcommand{\arraystretch}{1.0}
\end{table}

\begin{table}
\caption{\label{tablepar1} Fit parameters from the best fits of
the $\pi\pi$ mass spectrum in $e^+e^- \to J/\psi \pi^+\pi^-$ and the ratios
${\sigma(e^+e^- \to J/\psi K^+ K^-)}/{\sigma(e^+e^- \to J/\psi \pi^+\pi^-)}$  at
$E=4.23\GeV$ (Fit~Ia and Ib) and $E=4.26\GeV$
(Fit~IIa, IIb, IIc, and IId), respectively, using the DP $T$-matrix parametrization.}
\renewcommand{\arraystretch}{1.2}
\begin{center}
\begin{tabular}{l|cccccc}
\toprule
         & Fit~Ia, DP
         & Fit~Ib, DP
         & Fit~IIa, DP
         & Fit~IIb, DP
         & Fit~IIc, DP
         & Fit~IId, DP\\
\hline
$g_1~[\text{GeV}^{-1}]$   &    $ -0.29\pm 0.04$  &   $ 1.87\pm 0.13$   & $ 0.21\pm 0.04$ & $ -0.99\pm 0.11$ & $ 0.52\pm 0.02$& $ 0.20\pm 0.08$\\
$h_1~[\text{GeV}^{-1}]$   &    $ -0.29\pm 0.02$  &   $ -0.31\pm 0.06$  & $ -0.32\pm 0.02$  & $ 0.03\pm 0.04$ & $ 0.02\pm 0.01$& $ 0.09\pm 0.04$ \\
$g_8~[\text{GeV}^{-1}]$     &    0 (\text{fixed})   & $1.25\pm 0.11$  &    0 (\text{fixed}) & $ -1.18\pm 0.03$ &  0 (\text{fixed})& $ 1.01\pm 0.10$ \\
$h_8~[\text{GeV}^{-1}]$     &    0 (\text{fixed})   & $-1.96\pm 0.10$   &    0 (\text{fixed}) & $ 1.70\pm 0.18$ & 0 (\text{fixed})& $ -1.28\pm 0.08$\\
$C_{Y\Psi}^{Z_c}\times 10^{2}$  &    $ 0.7\pm 0.6$    &   $ 2.0\pm 0.8$   & $ 4.6\pm 0.3$ & $ 6.9\pm 0.3$ & 0 (\text{fixed})& 0 (\text{fixed})\\
$C_{Y\Psi}^{\text{loop}}~[\text{GeV}^{-3}]$  &    $ 4.5\pm 1.0$    &   $ 38.8\pm 2.5$   & $ 12.5\pm 0.8$ & $ -19.4\pm 2.1$ & 0 (\text{fixed})& 0 (\text{fixed})\\
\hline
 ${\chi^2}/{\rm d.o.f.}$ &  $\frac{405.1}{(44-4)}=10.13$  &  $\frac{102.1}{(44-6)}=2.69$    &  $\frac{182.7}{(46-4)}=4.35$ &  $\frac{63.9}{(46-6)}=1.60$ &  $\frac{428.9}{(46-2)}=9.75$  &  $\frac{148.2}{(46-4)}=3.53$    \\
\botrule
\end{tabular}
\end{center}
\renewcommand{\arraystretch}{1.0}
\end{table}

\begin{table}
\caption{\label{tablepar2} Fit parameters from the best fits of
the $\pi\pi$ mass spectrum in $e^+e^- \to J/\psi \pi^+\pi^-$ and the ratios
${\sigma(e^+e^- \to J/\psi K^+ K^-)}/{\sigma(e^+e^- \to J/\psi \pi^+\pi^-)}$  at
$E=4.23\GeV$ (Fit~Ia and Ib) and $E=4.26\GeV$ (Fit~IIa, IIb, IIc, and IId), respectively,
using the BO $T$-matrix.}
\renewcommand{\arraystretch}{1.2}
\begin{center}
\begin{tabular}{l|cccccc}
\toprule
         & Fit~Ia, BO
         & Fit~Ib, BO
         & Fit~IIa, BO
         & Fit~IIb, BO
         & Fit~IIc, BO
         & Fit~IId, BO\\
\hline
$g_1~[\text{GeV}^{-1}]$   &    $ -0.20\pm 0.04$  &   $ 1.34\pm 0.08$   & $ 0.30\pm 0.04$ & $ -1.24\pm 0.05$& $ 0.57\pm 0.02$& $ 0.32\pm 0.11$ \\
$h_1~[\text{GeV}^{-1}]$   &    $ -0.32\pm 0.02$  &   $ -0.07\pm 0.03$  & $ -0.35\pm 0.01$  & $ 0.02\pm 0.03$& $ -0.02\pm 0.01$& $ -0.01\pm 0.06$  \\
$g_8~[\text{GeV}^{-1}]$     &    0 (\text{fixed})   & $1.65\pm 0.15$  &    0 (\text{fixed}) & $ -1.31\pm 0.05$&  0 (\text{fixed})& $ 0.85\pm 0.12$  \\
$h_8~[\text{GeV}^{-1}]$     &    0 (\text{fixed})   & $-2.37\pm 0.02$   &    0 (\text{fixed}) & $ 2.03\pm 0.06$&  0 (\text{fixed})& $ -1.14\pm 0.11$ \\
$C_{Y\Psi}^{Z_c} \times 10^{2}$  &    $ 6.3\pm 0.6$ & $ 3.4\pm 0.7$& $  6.5\pm 0.2$ & $ 8.0\pm 0.2$ &  0 (\text{fixed})& 0 (\text{fixed})\\
$C_{Y\Psi}^{\text{loop}} ~[\text{GeV}^{-3}]$  &    $ 8.0\pm 0.8$    &   $ 40.9\pm 3.6$   & $8.7\pm 1.0$ & $-34.0\pm 1.9$ &  0 (\text{fixed})& 0 (\text{fixed})\\
\hline
 ${\chi^2}/{\rm d.o.f.}$ &  $\frac{308.7}{(44-4)}=7.72$  &  $\frac{121.4}{(44-6)}=3.19$    &  $\frac{170.4}{(46-4)}=4.06$ &  $\frac{94.3}{(46-6)}=2.36$ &  $\frac{446.5}{(46-2)}=10.15$&  $\frac{176.7}{(46-4)}=4.21$   \\
\botrule
\end{tabular}
\end{center}
\renewcommand{\arraystretch}{1.0}
\end{table}

The uncertainty due to the
dispersive input for the $\pi\pi/K\bar{K}$ rescattering is estimated by comparing the fits with the two different $T_0^0(s)$ matrices (DP~\cite{Dai:2014lza,Dai:2014zta,Dai:2016ytz} vs.\ BO~\cite{Leutwyler2012,Moussallam2004}).
In Fig.~\ref{fig.Mpipi}, the best fit results of the $\pi\pi$ mass spectrum in $e^+e^- \to J/\psi \pi^+\pi^-$
are shown, where the borders of the bands represent the fit results using these two different $T_0^0(s)$ matrix parametrizations. The fit results of the ratios of the cross sections ${\sigma(e^+e^- \to J/\psi K^+ K^-)}/{\sigma(e^+e^- \to J/\psi \pi^+\pi^-)}$ are given in Table~\ref{table-ratioes}. The fitted parameters as well as the $\chi^2/\text{d.o.f.}$
are shown in Tables~\ref{tablepar1} and~\ref{tablepar2} for the DP and BO parametrizations, respectively.
As can be seen from Fig.~\ref{fig.Mpipi} as well as Tables~\ref{tablepar1} and~\ref{tablepar2},
the fit quality to the data set at $E=4.23\GeV$ is
worse than that at $E=4.26\GeV$, in particular in the region close to the lower kinematical boundary and for the highest data point.
Notice that by using the inputs from known scattering observables in the dispersion relations, the effects of resonances in the considered partial waves, i.e.\ the $f_0(500)$, $f_0(980)$, and $f_2(1270)$, are included automatically.  Since the dataset at $E=4.26\GeV$ has a larger phase space to reveal the
nontrivial structure and the fits are better, we discuss the fit results of
this data set in more details.

\begin{figure}
\centering
\includegraphics[width=\linewidth]{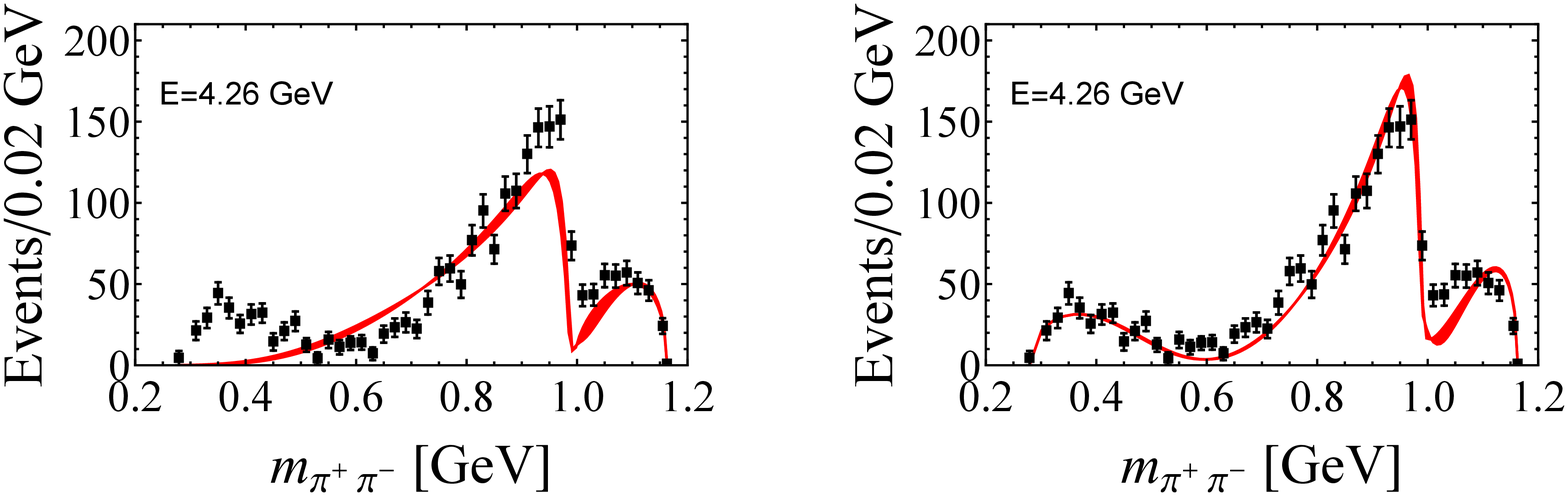}
\caption{Fit results of the $\pi\pi$ invariant mass spectra in
$e^+e^- \to J/\psi \pi^+\pi^-$ for Fits~IIc (left) and IId (right). The borders of the bands represent our best fit results using two different $T_0^0(s)$ matrices. The background-subtracted and efficiency-corrected experimental data are taken from Ref.~\cite{Collaboration:2017njt}. }\label{fig.Mpipi_4260_NoLefthandcuts}
\end{figure}

It is interesting to compare Fits~IIa and IIb. In Fit~IIa, the
SU(3) octet chiral contact terms are not included. The experimental
data, especially the broad peak in the region lower than $0.6\GeV$, cannot be described well.
In contrast, in Fit~IIb, including the SU(3) octet chiral
contact terms, the fit quality is improved significantly. A similar improvement is also observed comparing Fits~Ib and Ia.
We also perform two further Fits~IIc and IId for the $E=4.26\GeV$ dataset, considering only the contact terms and switching off the left-hand cuts: in Fit~IIc we only retain the SU(3) singlet component, while in Fit~IId, both the SU(3) singlet and octet components are taken into account. The result is shown in Fig.~\ref{fig.Mpipi_4260_NoLefthandcuts}, and the fit couplings are also listed in Table~\ref{tablepar1}. Comparing Fits~IIc and IId, one also finds that adding the SU(3) octet component increases the fit quality significantly.

It is instructive
to analyze the ratio of the parameters for the SU(3) octet component relative to those for
the SU(3) singlet component.
Using the results of Fit~IIb as shown in Tables~\ref{tablepar1} and~\ref{tablepar2}, we have $g_8/g_1=1.2 \pm 0.2$ and $h_8/h_1=57\pm 76$ in the DP parametrization and $g_8/g_1=1.1 \pm 0.1$ and $h_8/h_1=102\pm 152$ in the BO one,
which agree well with each other within errors. Note that $h_8/h_1$ is not as stable as $g_8/g_1$: the reason is that $h_1$ is small in most fits. In the $\bar{D}D_1$
hadronic molecule scenario of $Y(4260)$, one has
\be
|Y(4260)\rangle=\frac{1}{2}\big[|D_1^0 \bar{D}^0\rangle+|D_1^+
D^-\rangle\big]+\mathrm{c.c.}\,,
\ee
from which the light-quark component reads $|u \bar{u}+d\bar{d}\rangle/\sqrt{2} = (\sqrt{2}
V_1^{\text{light}} +V_8^{\text{light}})/\sqrt{3}$, where the definitions of
the singlet and octet components $V_1^{\text{light}}$
and $V_8^{\text{light}}$ have been given below
Eq.~\eqref{eq.YComponents}. They thus give the ratio of $1/\sqrt{2}$.
Certainly our results (values of $g_8/g_1$) differ significantly from the result of the pure $\bar{D}D_1$
hadronic molecule scenario. In addition to the
$\bar{D}D_1$ hadronic molecule, the $Y(4260)$ may contain other
SU(3) singlet sources, e.g., from $|c\bar{c}\rangle$ or a hybrid. Assuming in
the transition $Y\to\psi\Phi\Phi$ the strengths of the light-quark
components from the $\bar{D}D_1$ hadronic molecule and the other SU(3)
singlet source are $\alpha$ and $\beta$, respectively, namely,\footnote{Notice that any isoscalar pair of nonstrange charm and anticharm mesons has the same SU(3) structure.}
\be
\frac{\alpha}{\sqrt{3}}\Big(\sqrt{2}V_1^{\text{light}}
+V_8^{\text{light}}\Big)+\beta\, V_1^{\text{light}}\,,
\ee
we can
estimate the ratio of $\beta/\alpha=-0.30\pm 0.05$ based on our
results of $g_8/g_1$. Thus we conclude that there is a large
light-quark SU(3) octet component in the $Y(4260)$, and
scenarios of a hybrid or conventional charmonium are disfavored
since the light quarks have to be produced in the SU(3)
singlet state in such states. Also our study shows that the $\bar D D_1$ component of the $Y(4260)$ may not be completely dominant.
This is not unnatural, as the $Y(4260)$ mass, being around $4.22\GeV$, is about $70\MeV$ below the $\bar D D_1$ threshold.

\begin{figure}
\centering
\includegraphics[width=\linewidth]{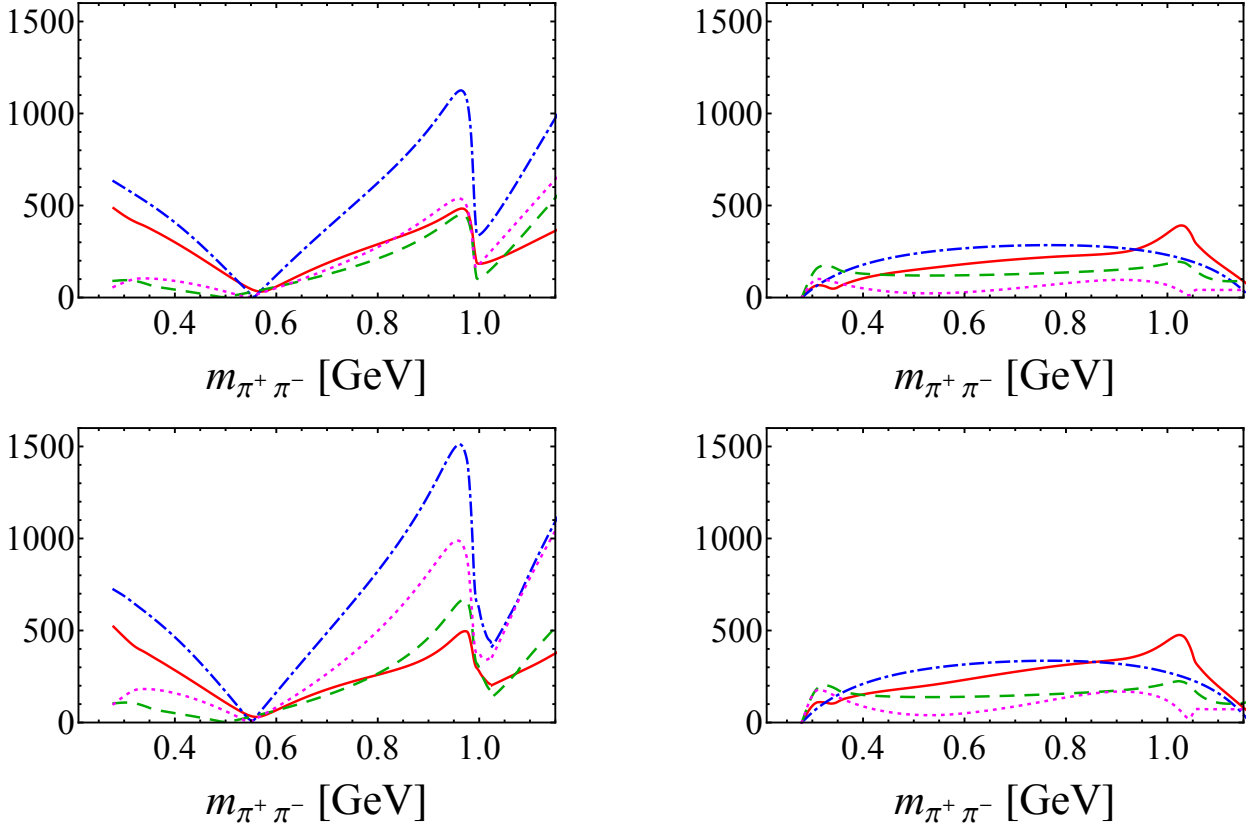}
\caption{The moduli of the $S$- (left) and $D$-wave (right)
amplitudes for $e^+e^- \to J/\psi \pi^+\pi^-$ in
Fit~IIb, using the DP (top) or the BO (bottom) parametrizations. The red solid lines represent our best fit results,
while the blue dot-dashed, darker green dashed,
and magenta dotted lines correspond to the contributions
from the chiral contact terms, $Z_c$-exchange, and the triangle diagrams, respectively.
}
\label{fig.Moduli}
\end{figure}
In Fig.~\ref{fig.Moduli}, we plot the moduli of the $S$- and
$D$-wave amplitudes from the chiral contact terms, the $Z_c$-exchange terms, and the triangle diagrams for Fit~IIb.
An interesting feature is that the $D$-wave contribution is comparable
to the $S$-wave contribution in almost the whole phase space. Such a large $D$-wave
contribution in the $Y\psi \Phi\Phi$ transition again indicates that the $Y(4260)$ cannot be a conventional charmonium state, for which the $\pi\pi$ $S$-wave should be dominant.
Notice that in the $\bar D D_1$ hadronic molecule interpretation~\cite{Cleven:2013mka,Lu:2017yhl}, the $\pi\pi$ $D$-wave emerges naturally since the $D_1$ decays dominantly into $D$-wave $D^*\pi$.
Also one observes that the contributions from the chiral contact terms and the l.h.c.\ contributions are of the same order. Amongst the l.h.c.\ contributions, both the $Z_c$ term and the triangle diagrams appear far from negligible. A better distinction of the effects of the
$Z_c$ and the open-charm loops requires a detailed analysis of the $J/\psi\pi$ distribution and is beyond the scope of the present paper.

\section{Conclusions}
\label{conclu}

We have used dispersion theory to study the processes
$e^+e^-\to Y(4260) \to J/\psi \pi\pi(K\bar{K})$. In particular, we have analyzed the roles of the
light-quark SU(3) singlet state and SU(3) octet state in these
transitions. The strong FSI, especially the coupled-channel ($\pi\pi$ and $K\bar{K}$) FSI in
the $S$-wave, has been considered in a model-independent way, and
the leading chiral amplitude acts as the subtraction function in the
modified Omn\`es solution.
Through fitting to the data of the $\pi\pi$
invariant mass spectra of $e^+e^-\to Y(4260) \to J/\psi \pi\pi$
and the ratios of the cross sections ${\sigma(e^+e^- \to J/\psi K^+ K^-)}/{\sigma(e^+e^- \to J/\psi \pi^+\pi^-)}$,
we find that the light-quark SU(3) octet state plays a
significant role in the $Y(4260)J/\psi\Phi\Phi$ transition, which
indicates that the $Y(4260)$ contains a large light-quark component.
Thus we conclude that the $Y(4260)$ is in all likelihood neither a hybrid nor a
conventional charmonium state. Furthermore, through an analysis of the ratio
of the light-quark SU(3) octet and singlet
components, we show that the $Y(4260)$ does not behave like a pure $\bar D D_1$ hadronic
molecule. We also find that there is a large $D$-wave component in the $\pi\pi$ invariant spectrum of the $Y(4260)$.
We close this manuscript by anticipating a combined analysis of both the $Y(4260)$ and $Z_c(3900)$ data. Such an analysis is a necessary step toward revealing the nature of both states, as there is evidence that the $Z_c(3900)$ events in the $J/\psi\pi\pi$ are only produced when the latter is constrained in the $Y(4260)$ region~\cite{Abazov:2018cyu}.

\section*{Acknowledgments}

We acknowledge Rong-Gang Ping for helpful discussions on the experimental analyses and for kindly providing us with the efficiency-corrected data, and Christoph Hanhart for
a careful reading of the manuscript and valuable comments and suggestions.
This work is supported in part by the Fundamental Research Funds
for the Central Universities under Grants No.~531107051122 and No.~06500077, by the National Natural Science Foundation of China (NSFC) under Grants No.~11805059, No.~11747601, and No.~11835015, by NSFC and Deutsche Forschungsgemeinschaft (DFG) through
funds provided to the Sino--German Collaborative Research Center ``Symmetries and the
Emergence of Structure in QCD'' (NSFC Grant No.~11621131001,
DFG Grant No.~TRR110), by the Thousand Talents Plan for Young
Professionals, by the CAS Key Research Program of Frontier Sciences
(Grant No.~QYZDB-SSW-SYS013), by the CAS Key Research Program (Grant No.~XDPB09), and by the CAS Center for Excellence in Particle Physics (CCEPP).


\begin{thebibliography}{99}

\bibitem{Aubert:2005rm}
  B.~Aubert {\it et al.} ({\it BABAR} Collaboration),
  Phys.\ Rev.\ Lett.\  {\bf 95}, 142001 (2005).

\bibitem{Godfrey}
  S.~Godfrey and N.~Isgur,
  Phys.\ Rev.\  D {\bf 32}, 189 (1985).

\bibitem{Pakhlova:2009jv}
  G.~Pakhlova {\it et al.} ({\it Belle} Collaboration),
  Phys.\ Rev.\ D {\bf 80}, 091101 (2009).

\bibitem{Chen:2016qju}
  H.-X.~Chen, W.~Chen, X.~Liu, and S.-L.~Zhu,
  Phys.\ Rep.\  {\bf 639}, 1 (2016).

\bibitem{Hosaka:2016pey}
  A.~Hosaka, T.~Iijima, K.~Miyabayashi, Y.~Sakai, and S.~Yasui,
  Prog.\  Theor.\ Exp.\ Phys.\ {\bf 2016}, 062C01 (2016).

\bibitem{Lebed:2016hpi}
  R.~F.~Lebed, R.~E.~Mitchell, and E.~S.~Swanson,
  Prog.\ Part.\ Nucl.\ Phys.\  {\bf 93}, 143 (2017).

\bibitem{Esposito:2016noz}
  A.~Esposito, A.~Pilloni, and A.~D.~Polosa,
  Phys.\ Rep.\  {\bf 668}, 1 (2016).

\bibitem{Guo:2017jvc}
  F.-K.~Guo, C.~Hanhart, U.-G.~Mei\ss ner, Q.~Wang, Q.~Zhao, and B.-S.~Zou,
  Rev.\ Mod.\ Phys.\  {\bf 90},  015004 (2018).

\bibitem{Ali:2017jda}
  A.~Ali, J.~S.~Lange, and S.~Stone,
  Prog.\ Part.\ Nucl.\ Phys.\  {\bf 97}, 123 (2017).

\bibitem{Olsen:2017bmm}
  S.~L.~Olsen, T.~Skwarnicki, and D.~Zieminska,
  Rev.\ Mod.\ Phys.\  {\bf 90}, 015003 (2018).

\bibitem{Karliner:2017qhf}
  M.~Karliner, J.~L.~Rosner, and T.~Skwarnicki,
  Ann.\ Rev.\ Nucl.\ Part.\ Sci.\  {\bf 68}, 17 (2018).

\bibitem{Yuan:2018inv}
  C.-Z.~Yuan,
  Int.\ J.\ Mod.\ Phys.\ A {\bf 33}, 1830018 (2018).

\bibitem{Kou:2018nap}
  E.~Kou {\it et al.},
  arXiv:1808.10567.

\bibitem{Cerri:2018ypt}
  A.~Cerri {\it et al.},
  arXiv:1812.07638.

\bibitem{Zhu:2005hp}
  S.-L.~Zhu,
  Phys.\ Lett.\ B {\bf 625}, 212 (2005).

\bibitem{Close:2005iz}
  F.~E.~Close and P.~R.~Page,
  Phys.\ Lett.\ B {\bf 628}, 215 (2005).

\bibitem{Kalashnikova:2008qr}
  Y.~S.~Kalashnikova and A.~V.~Nefediev,
  Phys.\ Rev.\ D {\bf 77}, 054025 (2008).

\bibitem{LlanesEstrada:2005hz}
  F.~J.~Llanes-Estrada,
  Phys.\ Rev.\ D {\bf 72}, 031503 (2005).

\bibitem{Li:2009zu}
  B.-Q.~Li and K.-T.~Chao,
  Phys.\ Rev.\ D {\bf 79}, 094004 (2009).

\bibitem{Shah:2012js}
  M.~Shah, A.~Parmar, and P.~C.~Vinodkumar,
  Phys.\ Rev.\ D {\bf 86}, 034015 (2012).

\bibitem{Qiao:2005av}
  C.-F.~Qiao,
  Phys.\ Lett.\ B {\bf 639}, 263 (2006).

\bibitem{Dubynskiy:2008mq}
  S.~Dubynskiy and M.~B.~Voloshin,
  Phys.\ Lett.\ B {\bf 666}, 344 (2008).

\bibitem{Li:2013ssa}
  X.~Li and M.~B.~Voloshin,
  Mod.\ Phys.\ Lett.\ A {\bf 29}, 1450060 (2014).

\bibitem{Maiani:2005pe}
  L.~Maiani, V.~Riquer, F.~Piccinini, and A.~D.~Polosa,
  Phys.\ Rev.\ D {\bf 72}, 031502 (2005).


\bibitem{Ali:2017wsf}
  A.~Ali, L.~Maiani, A.~V.~Borisov, I.~Ahmed, M.~Jamil Aslam, A.~Y.~Parkhomenko, A.~D.~Polosa and A.~Rehman,
  Eur.\ Phys.\ J.\ C {\bf 78}, 29 (2018).


\bibitem{Wang:2018ntv}
  Z.-G.~Wang,
  Eur.\ Phys.\ J.\ C {\bf 78}, 933 (2018).

\bibitem{Ding:2008gr}
  G.-J.~Ding,
  Phys.\ Rev.\ D {\bf 79}, 014001 (2009).

\bibitem{Wang:2013cya}
  Q.~Wang, C.~Hanhart, and Q.~Zhao,
  Phys.\ Rev.\ Lett.\  {\bf 111}, 132003 (2013).

\bibitem{Li:2013yla}
  G.~Li and X.-H.~Liu,
  Phys.\ Rev.\ D {\bf 88}, 094008 (2013).

\bibitem{Cleven:2013mka}
  M.~Cleven, Q.~Wang, F.-K.~Guo, C.~Hanhart, U.-G.~Mei\ss ner, and Q.~Zhao,
  Phys.\ Rev.\ D {\bf 90},  074039 (2014).

\bibitem{Dai:2012pb}
  L.-Y.~Dai, M.~Shi, G.-Y.~Tang, and H.-Q.~Zheng,
  Phys.\ Rev.\ D {\bf 92}, 014020 (2015).

\bibitem{Chen:2010nv}
  D.-Y.~Chen, J.~He, and X.~Liu,
  Phys.\ Rev.\ D {\bf 83}, 054021 (2011).

\bibitem{Chen:2017uof}
  D.~Y.~Chen, X.~Liu, and T.~Matsuki,
  Eur.\ Phys.\ J.\ C {\bf 78}, 136 (2018).

\bibitem{Ablikim:2016qzw}
  M.~Ablikim {\it et al.} (BESIII Collaboration),
  Phys.\ Rev.\ Lett.\  {\bf 118}, 092001 (2017).

\bibitem{BESIII:2016adj}
  M.~Ablikim {\it et al.} (BESIII Collaboration),
  Phys.\ Rev.\ Lett.\  {\bf 118}, 092002 (2017).

\bibitem{Ablikim:2015uix}
  M.~Ablikim {\it et al.} (BESIII Collaboration),
  Phys.\ Rev.\ D {\bf 93}, 011102 (2016).

\bibitem{Ablikim:2013dyn}
  M.~Ablikim {\it et al.} (BESIII Collaboration),
  Phys.\ Rev.\ Lett.\  {\bf 112}, 092001 (2014).

\bibitem{Ablikim:2017oaf}
  M.~Ablikim {\it et al.} (BESIII Collaboration),
  Phys.\ Rev.\ D {\bf 96}, 032004 (2017);
  Phys.\ Rev.\ D {\bf 99}, 019903(E) (2019).

\bibitem{Ablikim:2018vxx}
  M.~Ablikim {\it et al.} (BESIII Collaboration),
 Phys.\ Rev.\ Lett.\  {\bf 122}, 102002 (2019).

\bibitem{Qin:2016spb}
  W.~Qin, S.-R.~Xue, and Q.~Zhao,
  Phys.\ Rev.\ D {\bf 94}, 054035 (2016).

\bibitem{Wang:2013hga}
  Q.~Wang, C.~Hanhart, and Q.~Zhao,
  Phys.\ Lett.\ B {\bf 725}, 106 (2013).

\bibitem{Albaladejo:2015lob}
  M.~Albaladejo, F.-K.~Guo, C.~Hidalgo-Duque, and J.~Nieves,
  Phys.\ Lett.\ B {\bf 755}, 337 (2016).

\bibitem{Mannel}
  T.~Mannel and R.~Urech,
  Z.\ Phys.\ C {\bf 73}, 541 (1997).

\bibitem{Chen2016}
  Y.-H.~Chen, J.~T.~Daub, F.-K.~Guo, B.~Kubis, U.-G.~Mei\ss ner, and
  B.-S.~Zou,
  Phys.\ Rev.\ D {\bf 93}, 034030 (2016).

\bibitem{Chen:2016mjn}
  Y.-H.~Chen, M.~Cleven, J.~T.~Daub, F.-K.~Guo, C.~Hanhart, B.~Kubis, U.-G.~Mei\ss ner, and B.-S.~Zou,
  Phys.\ Rev.\ D {\bf 95}, 034022 (2017).

\bibitem{Shen:2014gdm}
  C.~P.~Shen {\it et al.} ({\it Belle} Collaboration),
  Phys.\ Rev.\ D {\bf 89}, 072015 (2014).

\bibitem{Guo2011}
  M.~Cleven, F.-K.~Guo, C.~Hanhart, and U.-G.~Mei\ss ner,
  Eur.\ Phys.\ J.\ A {\bf 47}, 120 (2011).

\bibitem{Colangelo:2005gb}
  P.~Colangelo, F.~De Fazio, and R.~Ferrandes,
  Phys.\ Lett.\ B {\bf 634}, 235 (2006).

\bibitem{Guo:2013nza}
  F.-K.~Guo, C.~Hanhart, U.-G.~Mei\ss ner, Q.~Wang, and Q.~Zhao,
  Phys.\ Lett.\ B {\bf 725}, 127 (2013).

\bibitem{Mehen2013}
  T.~Mehen and J.~W.~Powell,
  Phys.\ Rev.\ D {\bf 88}, 034017 (2013).

\bibitem{Mehen2008}
  S.~Fleming and T.~Mehen,
  Phys.\ Rev.\ D {\bf 78}, 094019 (2008).

\bibitem{Moussallam:2013una}
  B.~Moussallam,
  Eur.\ Phys.\ J.\ C {\bf 73}, 2539 (2013).

\bibitem{Schmid:1967ojm}
  C.~Schmid,
  Phys.\ Rev.\  {\bf 154}, 1363 (1967).

\bibitem{Hanhart:2015cua}
  C.~Hanhart, Y.~S.~Kalashnikova, P.~Matuschek, R.~V.~Mizuk, A.~V.~Nefediev, and Q.~Wang,
  Phys.\ Rev.\ Lett.\  {\bf 115}, 202001 (2015).

\bibitem{Guo:2016bjq}
  F.-K.~Guo, C.~Hanhart, Y.~S.~Kalashnikova, P.~Matuschek, R.~V.~Mizuk, A.~V.~Nefediev, Q.~Wang, and J.-L.~Wynen,
  Phys.\ Rev.\ D {\bf 93}, 074031 (2016).

\bibitem{Pilloni:2016obd}
  A.~Pilloni {\it et al.} (JPAC Collaboration),
  Phys.\ Lett.\ B {\bf 772}, 200 (2017).

\bibitem{Barton:1961ms}
  G.~Barton and C.~Kacser,
  Nuovo Cimento\  {\bf 21}, 593 (1961).

\bibitem{Bronzan:1963mby}
  J.~B.~Bronzan and C.~Kacser,
  Phys.\ Rev.\  {\bf 132}, 2703 (1963).

\bibitem{Kambor:1995yc}
  J.~Kambor, C.~Wiesendanger, and D.~Wyler,
  Nucl.\ Phys.\  {\bf B465}, 215 (1996).

\bibitem{Moussallam-gamma}
  R.~Garc\'ia-Mart\'in and B.~Moussallam,
  Eur.\ Phys.\ J.\ C {\bf 70}, 155 (2010).

\bibitem{KubisPlenter}
  B.~Kubis and J.~Plenter,
  Eur.\ Phys.\ J.\ C {\bf 75}, 283 (2015).

\bibitem{ZHGuo}
  Z.-H.~Guo and J.~A.~Oller,
  Phys.\ Rev.\ D {\bf 84}, 034005 (2011).

\bibitem{Kang}
  X.-W.~Kang, B.~Kubis, C.~Hanhart, and U.-G.~Mei\ss ner,
  Phys.\ Rev.\ D {\bf 89}, 053015 (2014).

\bibitem{Dai:2014lza}
  L.-Y.~Dai and M.~R.~Pennington,
  Phys.\ Lett.\ B {\bf 736}, 11 (2014).

\bibitem{Dai:2014zta}
  L.-Y.~Dai and M.~R.~Pennington,
  Phys.\ Rev.\ D {\bf 90}, 036004 (2014).

\bibitem{Dai:2016ytz}
  L.-Y.~Dai and M.~R.~Pennington,
  Phys.\ Rev.\ D {\bf 94}, 116021 (2016).

\bibitem{Leutwyler2012}
  I.~Caprini, G.~Colangelo, and H.~Leutwyler,
  Eur.\ Phys.\ J.\ C {\bf 72}, 1860 (2012).

\bibitem{Moussallam2004}
  P.~B\"uttiker, S.~Descotes-Genon, and B.~Moussallam,
  Eur.\ Phys.\ J.\ C {\bf 33}, 409 (2004).

\bibitem{Tanabashi:2018oca}
  M.~Tanabashi {\it et al.} [Particle Data Group],
  Phys.\ Rev.\ D {\bf 98}, 030001 (2018).

\bibitem{Ropertz:2018stk}
  S.~Ropertz, C.~Hanhart, and B.~Kubis,
  Eur.\ Phys.\ J.\ C {\bf 78}, 1000 (2018).

\bibitem{Moussallam2000}
  B.~Moussallam,
  Eur.\ Phys.\ J.\ C {\bf 14}, 111 (2000).

\bibitem{Leutwyler90}
  J.~F.~Donoghue, J.~Gasser, and H.~Leutwyler,
  Nucl.\ Phys.\  {\bf B343}, 341 (1990).

\bibitem{Hoferichter:2012wf}
  M.~Hoferichter, C.~Ditsche, B.~Kubis, and U.-G.~Mei{\ss}ner,
  J.~High Energy Phys.\ {\bf 06} (2012) 063.

\bibitem{Daub}
  J.~T.~Daub, C.~Hanhart, and B.~Kubis,
  J.~High Energy Phys.\ {\bf 02} (2016) 009.

\bibitem{Watson1}
  K.~M.~Watson,
  Phys.\ Rev.\  {\bf 88}, 1163 (1952).

\bibitem{Watson2}
  K.~M.~Watson,
  Phys.\ Rev.\  {\bf 95}, 228 (1954).

\bibitem{Leutwyler96}
  A.~V.~Anisovich and H.~Leutwyler,
  Phys.\ Lett.\ B {\bf 375}, 335 (1996).

\bibitem{Omnes}
  R.~Omn\`es,
  Nuovo Cimento\  {\bf 8}, 316 (1958).

\bibitem{Pelaez}
  R.~Garc\'ia-Mart\'in, R.~Kami\'nski, J.~R.~Pel\'aez, J.~Ruiz de Elvira, and F.~J.~Yndur\'ain,
  Phys.\ Rev.\  D {\bf 83}, 074004 (2011).

\bibitem{Daub:2012mu}
  J.~T.~Daub, H.~K.~Dreiner, C.~Hanhart, B.~Kubis, and U.-G.~Mei\ss ner,
  J.~High Energy Phys.\ {\bf 01} (2013) 179.

\bibitem{Collaboration:2017njt}
  M.~Ablikim {\it et al.} (BESIII Collaboration),
  Phys.\ Rev.\ Lett.\  {\bf 119}, 072001 (2017).

\bibitem{Ablikim:2018epj}
  M.~Ablikim {\it et al.} (BESIII Collaboration),
  Phys.\ Rev.\ D {\bf 97}, 071101 (2018).

\bibitem{Lu:2017yhl}
Y.~Lu, M.~N.~Anwar, and B.-S.~Zou,
Phys.\ Rev.\ D {\bf 96}, 114022 (2017).

\bibitem{Abazov:2018cyu}
  V.~M.~Abazov {\it et al.} (D0 Collaboration),
  Phys.\ Rev.\ D {\bf 98}, 052010 (2018).


\end{thebibliography}
\end{document}